\documentclass[aps]{revtex4}
\usepackage{graphicx,psfrag}
\usepackage{color}
\usepackage{graphicx}

\def\bc{\begin{center}}
\def\ec{\end{center}}

\def\beq{\begin{equation}}
\def\eeq{\end{equation}}

\def\bk{{\bf k}}

\def\bc{\begin{center}}
\def\ec{\end{center}}

\def\beq{\begin{equation}}
\def\eeq{\end{equation}}

\def\bk{{\bf k}}

\begin{document}

\title{Electron-hole superfluidity controlled by a periodic
potential}
\author{Oleg L. Berman$^{1,2}$, Roman Ya. Kezerashvili$^{1,2}$, Yurii E.
Lozovik$^{3,4}$, and Klaus G. Ziegler$^{5}$}
\affiliation{\mbox{$^{1}$Physics Department, New York City College
of Technology, The City University of New York,} \\
Brooklyn, NY 11201, USA \\
\mbox{$^{2}$The Graduate School and University Center, The
City University of New York,} \\
New York, NY 10016, USA \\
\mbox{$^{3}$Institute of Spectroscopy, Russian Academy of
Sciences,} \\
142190 Troitsk, Moscow, Russia \\
\mbox{$^{4}$Research University Higher School of  Economics, Moscow, Russia 101000}\\
\mbox{$^{5}$Institut f\"ur Physik, Universit\"at Augsburg\\
D-86135 Augsburg, Germany }}
\date{\today}

\begin{abstract}
We propose to control of an electron-hole superfluid in
semiconductor coupled quantum wells and double layers
of two-dimensional (2D) material by an external periodic field.
This can either be created by the gates periodically located and
attached to the quantum wells or double layers of 2D
material or by the Moir\'e pattern of two twisted layers. The
dependence of the electron-hole pairing order parameter on the
temperature, the charge carrier density, and the gate parameters is
obtained by minimization of the mean-field free energy. The second
order phase transition between superfluid and electron-hole plasma,
controlled by the external periodic gate field, is analyzed for
different parameters.
\end{abstract}

\maketitle

\section{Introduction}
\label{intro}

The system of spatially separated electrons and holes can be
realized in semiconductor coupled quantum wells (CQWs), where
electrons and holes are located in different quantum wells. For
low temperatures and weak attraction the
Bardeen-Cooper-Schrieffer (BCS) approach describes the superfluid
formed by coherent Cooper pairs, while in the strong attraction
regime the composite bosons, known as indirect (dipolar) excitons,
are formed.  An electron-hole plasma (EHP) appears at sufficiently
higher temperatures. Superfluidity in the two-dimensional (2D)
system with spatially separated
electrons and holes was predicted using the BCS mean-field approach~\cite%
{Lozovik}, which caused intensive theoretical~\cite%
{Shevchenko,Littlewood,Vignale,Ulloa,
DasSarma,Perali,Peeters,combescot17,Fil,usp}
as well as experimental studies~\cite%
{Zrenner,Sivan,Snoke,Chemla,Butovr,Timofeev,Krivolapchuk,Snoke_paper,Snoke_paper_Sc,Dubin_PRL,Dean_Nature}.
Different electron-hole phases, characterized by unique collective
properties, have been analyzed in the system of spatially separated
electrons and holes~\cite{LB}. The BCS phase of electron-hole Cooper
pairs in a dense electron-hole system~\cite{Lozovik} and a dilute
gas of indirect excitons,
formed as bound states of electron-hole pairs, were also analyzed in CQWs~%
\cite{BLSC}. Superfluidity of the BCS phase, formed by
spatially separated electrons and holes, can be manifested as
non-dissipative electric currents and quasi-Josephson
phenomena~\cite{Lozovik,Shevchenko}. Besides the superfluid phase a
Wigner supersolid state caused by dipolar repulsion in electron-hole
double layers was described~\cite{LBW,As1,As2,JBD}. The recent
theoretical and experimental achievements in the studies of the
superfluid dipolar exciton phases in CQWs were reviewed in Ref.~\cite%
{Snoke_review}. Probing the ground state of an electron-hole double
layer by low temperature transport was experimentally
performed~\cite{Pepper_review}, and the various experimental studies
of excitonic phases in CQWs were described in Ref.~\cite{Butov_JPCM}.

Another physical realization for indirect excitons, formed in an
electron-hole double layer, is a wide single GaAs/AlGaAs quantum
well with a finite width~\cite{Kukushkin}. In a wide single QW,
the transverse electric field separates electrons and holes at the
different boundaries of the QW~\cite{Kukushkin}. The advantage of a
wide single QW compared with CQWs is the smaller number of the QW
boundaries, which leads to the increase of the electron mobility.
Based on the photoluminescence pattern caused by electron-hole
recombination, the evidence for a condensate of indirect excitons,
electrically polarized in a GaAs wide single QW, was achieved
experimentally recently for the thickness of $15\ \mathrm{nm}$ of
the quantum well at the temperature $T = 370 \ \mathrm{mK}$
~\cite{Dubin}. A spontaneous condensation of trapped two-dimensional
dipolar excitons from an interacting gas into a dense liquid state
was observed in GaAs/AlGaAs CQWs with an interwell separation $D = 4 \ \mathrm{nm}$
at the temperatures  below a critical temperature $T_{c} \approx 4.8
\ \mathrm{K}$~\cite{Rapaport}. The transport of indirect excitons
with an interwell separation $D = 4 \ \mathrm{nm}$ in GaAs/AlGaAs
CQWs in linear lattices, created by laterally modulated gate voltage
with a lattice period $b = 2 \ \mathrm{\mu m}$, was studied
experimentally at the temperatures $T = 1.6 \ \mathrm{K}$ and $T = 6
\ \mathrm{K}$, and the localization-delocalization transition for
transport across the lattice was observed with reducing lattice
amplitude or increasing exciton density~\cite{Butov_per}.

Besides semiconductor CQWs, the superfluid system of
spatially separated electrons and holes can appear in a graphene
double layer~\cite{BLG,Sokolik,Bist,BKZg,Perali}, two opposite
surfaces of the film of topological insulators~\cite{Efimkin}, two
layers with composite fermions in the quantum Hall regime at the
filling factor $\nu = 1/2$~\cite{EM}.

Today an intriguing counterpart to gapless graphene is a class of
monolayer direct bandgap materials, namely transition metal
dichalcogenides (TMDCs). Monolayers of TMDC  such as
$\mathrm{Mo S_{2}}$, $\mathrm{Mo Se_{2}}$, $\mathrm{Mo Te_{2}}$, $\mathrm{W S_{2}}$, $%
\mathrm{W Se_{2}}$, and $\mathrm{W Te_{2}}$ are 2D semiconductors,
which have a variety of applications in electronics and
opto-electronics~\cite{Kormanyos}. The strong interest in TMDC
monolayers is  motivated by the following facts: a semiconductor
band structure is characterized by a direct gap in the single
particle spectrum~\cite{Mak2010}, the existence of excitonic valley
physics, and the possibility of an electrically tunable, strong
light-matter interactions~\cite{Xiao,Mak2013}. Monolayers of
transition metal dichalcogenides are truly  2D semiconductors, which
hold great appeal for electronics and opto-electronics applications
due to their direct band gap properties. Monolayer TMDCs have
already been implemented in field-effect transistors, logical
devices, and lateral and tunneling optoelectronic
structures~\cite{Kormanyos}. Like graphene, the  monolayer TMDCs
have hexagonal lattice structures, and the extrema (valleys) in the
dispersion relations of both the valence and conduction bands can be
found at the $\mathbf{K}$ and $\mathbf{-K}$ points of the hexagonal Brillouin
zone. However, unlike graphene, these 2D crystals do not have
inversion symmetry~\cite{Kormanyos}.

High-temperature superfluidity can be studied in a heterostructure
of two TMDC monolayers, separated by a hexagonal boron nitride
($h$-BN) insulating barrier~\cite{Fogler}.
The dipolar excitons were observed in heterostructures formed by monolayers of $\mathrm{%
Mo S_{2}}$ on a substrate constrained by hexagonal boron nitride
layers~\cite{Calman}. The theoretical study of the phase diagram of 2D
condensates of indirect excitons in a TMDC double layer was reported~\cite{MacDonald_TMDC}.
The high-temperature superfluidity of the two component Bose gas of A
and B dipolar excitons in a transition metal dichalcogenide double
layer was predicted in Refs.~[\onlinecite{BK,BK2}].

In this paper we study how the BCS-like EHP-superfluid phase
transition can be controlled by the external periodic field, applied
to the spatially separated electrons and holes via the gates
periodically attached to CQWs, where a
quasi-two-dimensional system of charge carriers is formed. The
external periodic field, applied to the spatially separated
electrons and holes can be also produced via the gates periodically
attached to double layers of 2D material or a twisted TMDC double
layer, where a truly 2D system of charge carriers is formed. For
this purpose we employ a mean-field approximation for the many-body
system of electrons and holes, using the partition function of the
grand-canonical ensemble at the temperature $T$ and the chemical
potentials of the electrons and the holes, respectively. The latter
represent the Fermi energies of the electrons and the holes. The
logarithm of the partition function gives us immediately the free
energy, whose minimum defines the mean-field solution with a
non-vanishing order parameter of the superfluid phase. We also
briefly discuss quantum fluctuations around the mean-field solution
and how to measure them in terms of the static structure factor. The
main goal, though, is to analyze the influence of the external
periodic field on the critical temperature of the EHP-superfluid
transition in an electron-hole double layer.

The paper is organized in the following way. We obtain the free
energy of the electron-hole double layer in the external periodic
potential and study the second order
EHP-superfluid transition using a Landau expansion of the free energy in
Sec.~\ref{phase}. The results of calculations are presented and
analyzed in Sec.~\ref{res}. Finally, the discussion of the results
and the conclusions follow in Sec.~\ref{disc}.

\section{Phase transition in the electron-hole double layer under the action of the external
periodic potential}
\label{phase}

The Hamiltonian of the system of spatially separated electrons and
holes in momentum representation can be written as
\begin{equation}
H=\sum_{\mathbf{p}}\sum_{\sigma=e,h}\varepsilon_{\mathbf{p}%
,\sigma}c^\dagger_{\mathbf{p}\sigma}c_{\mathbf{p}\sigma} +\sum_{%
\mathbf{p},\mathbf{p}_{1},\mathbf{p}_{2}} U_\mathbf{p} c^\dagger_{\mathbf{p}-%
\mathbf{p}_{1},h}c_{\mathbf{p}-\mathbf{p}_{2},h}c^\dagger_{\mathbf{p}%
_{1},e}c_{\mathbf{p}_{2},e} \ ,  \label{hamilton}
\end{equation}
where $c^\dagger_{p,e}$ ($c_{p,e}$) is the creation (annihilation)
operator for electrons, and $c^\dagger_{p,h}$ ($c_{p,h}$) is the
corresponding operator for holes, The electron and hole
single-particle energy spectra $\varepsilon_{\mathbf{p},\sigma}$ are
defined below in the tight-binding approximation, reflecting the
external periodic field applied to the CQWs or double
layers of 2D material. The electron-hole attraction potential in
momentum space $U_\mathbf{p}$ is defined below. In
Eq.~(\ref{hamilton}) the spins of electrons and holes are neglected,
because we are not interested in magnetization effects.

\begin{figure}[tbp]
\includegraphics[width=10.0cm]{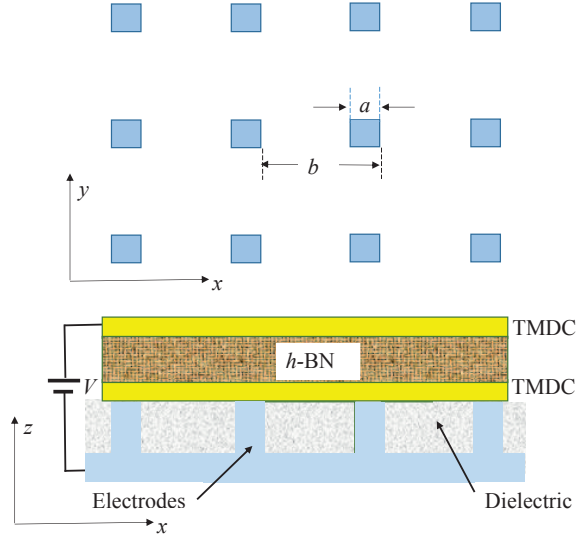}
\vspace{-5.00cm}
  \caption{Schematic electrode pattern in the $x-y$ and $z-x$ planes.}
  \label{scheme}
\end{figure}

We consider an external periodic potential induced by the gate
$V(\mathbf{r})$ forming either a 1D or a 2D square superlattice with
the period $b$ applied to the electron and hole quantum wells. As an
example, the particular case related to the phase transition of
indirect excitons in a double layer, formed by two
TMDC monolayers that are separated by \textit{h}-BN, since
\textit{h}-BN monolayers are characterized by relatively small
density of the defects of their crystal structure monolayers. In
Fig.~\ref{scheme} a schematic electrode pattern in the $x-y$ and
$z-x$ planes is presented. In our calculations we consider the TMDC
monolayers to be separated by \textit{h}-BN insulating layers and
the separation between two layers of TMDC materials calculated in
steps of $D_{hBN} = 0.333 \ \mathrm{nm}$, corresponding to the
thickness of one \textit{h}-BN monolayer~\cite{Fogler}. Therefore,
the interlayer separation $D$ is presented as $D$ = $N_{L}D_{hBN}$,
where $N_{L}$ is the number of \textit{h}-BN monolayers, placed
between two TMDC monolayers. It is obvious that the strength of the
electron-hole interaction decreases with the increase of the
separation between the layers. We assume that the densities of
electrons and holes are equal in order to have a neutral
electron-hole plasma and because the electrons and holes are created
always pairwise by an external laser source. This implies that the
corresponding chemical potentials are also equal. Periodically
positioned gates under the same electric potentials create in a turn
the periodical potential in the 2D system under consideration.

When the electron-hole attraction leads to Cooper
pairing of electrons and holes, characterized by the order parameter
$\Delta$ \cite{Schrieffer,DeGennes}, the free energy  is described
within the mean-field approximation (MFA)~\cite{Lozovik,BKZ}.
Following Ref.~\cite{BKZ}, and assuming that (i) the order parameter
$\Delta$ does not depend on the momentum, and (ii) the dispersion relation $\varepsilon _{\mathbf{p,}e}=\varepsilon _{\mathbf{p,}%
h}=\varepsilon _{\mathbf{p}}$ 
is the same for electrons and holes, the MFA of the free energy at temperature $T$
($\beta=1/k_B T$, where $k_B$ is the Boltzmann constant) as a function of the dimensionless order parameter
$\gamma=\beta|\Delta|$ can be written as
\begin{eqnarray}
F= 
-\frac{1}{\beta^2u_0}
\gamma^2 -\frac{1}{|B|\beta}\int_{B} \ln\left[%
2\left(1+ \cosh\left(\sqrt{\beta^2\varepsilon_{\mathbf{p}}^2 +\gamma^2}%
\right)\right)\right]d^2p \ .  \label{mfa0}
\end{eqnarray}
In Eq. (2) the integration over the momentum $\mathbf{p}$ is taken over the Brillouin zone with area $|B|$ and $u_0$ is the strength of the electron-hole interaction given by
\begin{eqnarray}  \label{ehint}
\frac{1}{u_{0}} = \frac{1}{|B|} \int_{B}
\frac{1}{U_\mathbf{p}}d^{2}p<0  ,
\end{eqnarray}
which parametrically depends on the inter layer separation $D$.

Let us rewrite Eq. (2) in the form of a dimensionless free energy $f
= - u_{0}F/(k_{B}T)^2$ and expand the latter in terms of the order
parameter $\gamma^2$ as $f=f_{0}+f_{2}\gamma ^{2}+f_{4}\gamma ^{4}$
(Landau expansion \cite{Landau, landau}). The corresponding
coefficients for this expansion are given in Appendix \ref{app:A} by
Eqs. (\ref{f0}) - (\ref{f4}). At the point of the phase transition
we have zero order parameter $\gamma = 0$, and the condition for the
minimum of the free energy is $\left.\frac{\partial f}{\partial
\gamma ^{2}} \right|_{\gamma = 0} = f_{2} = 0$. Therefore, the
critical point is defined by a vanishing coefficient $f_2$. Thus
from expression (\ref{f2}) follows that the critical inverse
temperature $\beta_c$ satisfies the condition \beq
\frac{1}{u_0}=-\frac{1}{|B|}\int_B\frac{\tanh(\beta_c|\epsilon_{\bf
p}|)}{|\epsilon_{\bf p}|}d^2p
=-\int_{E_0}^{E_1}\frac{\tanh(\beta_c|2tE-\delta_0|)}{|2tE-\delta_0|}\rho(E)dE
\ .
\label{mfa_1}
\eeq
Here we have used the fact that the $p$ integration can be expressed as an energy integration through
the relation $d^2p/|B|=\rho(E)dE$, where $\rho(E)$ is the density of states of the non-interacting Hamiltonian
with dispersion $\varepsilon_{\mathbf{p}}$. The dimensionless energy parameter
$E=(\delta_0-\varepsilon_{\mathbf{p}})/2t$ is derived from the dispersion which is shifted by the Fermi
energy $\delta_0$.
The integration is restricted to the interval $[E_0,E_1]$,
since only electronic states are accessible within the main band of the electronic band structure.
The specific values depend on the material and its dispersion, typically for a parabolic dispersion
they are given by $E_0\approx 0$ and $E_1\approx\hbar^2/(\lambda^22m)$ with the lattice constant
$\lambda$ of the underlying material.

The relation between $u_0$ and $\beta_c$ in Eq. (\ref{mfa_1}) indicates that an increasing
interaction strength $-u_0$ implies an increasing critical temperature. Moreover,
$\tanh(\beta_c|2tE-\delta_0|)/|2tE-\delta_0|$ is a monotonically decreasing function of $|2tE-\delta_0|$
with the maximum at $E=\delta_0/2t$. The density of states $\rho(E)$, on the other hand, distinguishes between
the case of a parabolic dispersion ($\rho(E)=const.$) and the case of a periodic potential,
where $\rho(E)$ is not constant. Thus the goal is to design the dispersion
by adding a superstructure to the material. This can either be achieved by doping~\cite{lee18}, by creating a
gated periodic potential on the 2D material~\cite{Butov_per} (cf. Fig. \ref{scheme}) or by twisting the two layers relative
to each other to create a Moir\'e pattern~\cite{morell10,bistritzer11}. Then the density of states can be chosen such
that it picks up the maximum value of the integrand
$\tanh(\beta_c|2tE-\delta_0|)/|2tE-\delta_0|$. Although this depends on the Fermi
energy $\delta_0$, the latter can also be tuned by a uniform external gate to obtain a large value
for the integral. As an example we consider the tight-binding approximation on a square lattice
with periodicity $b$.
Then the electron and the hole dispersion reads~\cite{Ziman,Simon,Ashcroft}:
\begin{eqnarray}
\varepsilon_{\mathbf{p}} = \delta_{0} -
2t\cos\left(p_{x}b/\hbar\right) - 2t\cos\left(p_{y}b/\hbar\right) \
\ \ (-\pi\le p_{x,y}b/\hbar <\pi) \ , \label{dispersion}
\end{eqnarray}
which has a band width $8t$. The corresponding density of states
reads~\cite{gonis}
\begin{eqnarray}  \label{Kfun}
\rho(E)=\rho_0\frac{K\left(\frac{2-|E|}{2+|E|}\right) }{2+|E|} \ \ \
(-2\le E\le 2) \ ,
\end{eqnarray}
where $K(x)$ is the complete elliptic integral of the first kind and
$\rho_0$ is a normalization factor. In Appendix B are given the expressions for the coefficients $f_{0},$ $f_{2}$ and
$f_{4}$  of the Landau expansion in the case of the periodical potential. For the dispersion
(\ref{dispersion}) we have derived some results directly from Eqs.
(\ref{mfa0}) and (\ref{mfa_1}), which will be discussed in the next
section.

Instead of the 2D periodic potential we could also consider an
anisotropic potential with a 1D periodicity, which would also affect
strongly the density of states. This case corresponds to the system
studied in Ref.~\cite{Butov_per}.

\section{Results}
\label{res}

\begin{figure}[tbp]
\includegraphics[width=17.0cm]{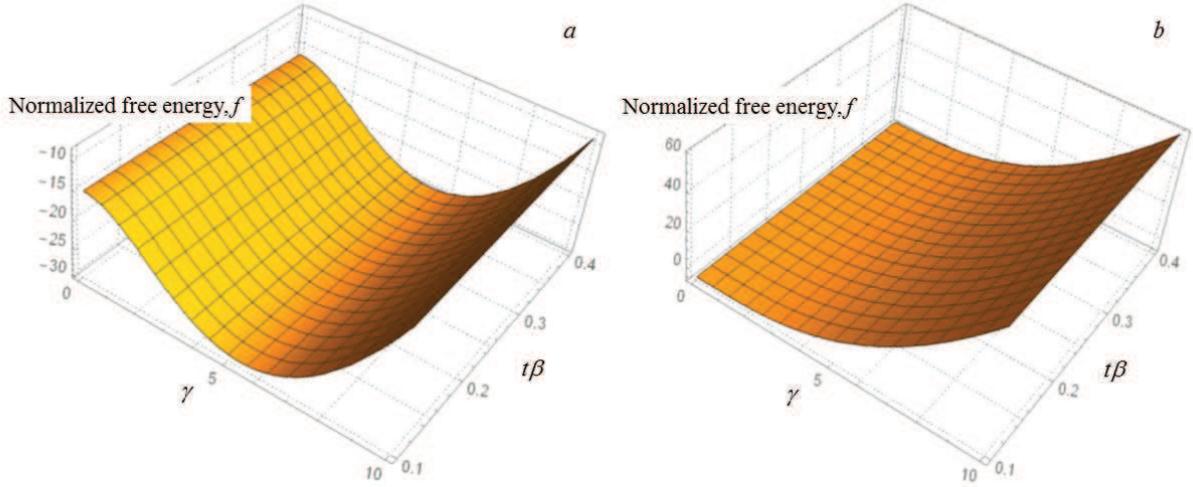}
\vspace{-2.5cm}
 \caption{The normalized free energy $f$ as a function of $t\beta$ and the order parameter
 $\gamma$. (a) $\delta_{0}/2t = 0.15$;  $\beta u_{0} = -80$.
 (b) $\delta_{0}/2t = 0.15$;  $\beta u_{0} = -30$. } \label{Fig_3D}
\end{figure}

\begin{figure}[tbp]
\includegraphics[width=20.0cm]{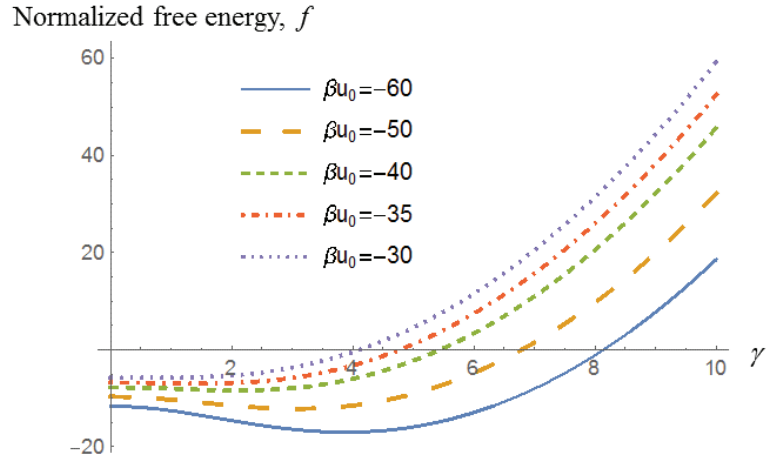}
\vspace{-2.5cm}
 \caption{The normalized free energy $f$ for $\beta u_{0} =
 -30$, $-35$, $-40$, $-50$, $-60$, at $\beta t = 0.22$ and $\delta_{0}/2t = 0.175$
 indicates a second order
phase transition between $\beta u_{0} = -50$ and $\beta u_{0} =
-40$. } \label{Fig2}
\end{figure}

The free energy $F$ of Eq.~(\ref{mfa0}) with the dispersion in Eq. (\ref{dispersion})
is calculated as a function of the dimensionless order parameter $\gamma$.
The dependence of the dimensionless free energy $f= - \beta^2u_{0}F$ on $t\beta$ and the
order parameter $\gamma$ is shown in Fig.~\ref{Fig_3D}. This result
demonstrates that $f$ has a minimum with
respect to the order parameter $\gamma$, while the dependence of $f$
on $t\beta$ is not strong. The non-zero minimum of $f$ with respect
to the order parameter $\gamma$ corresponds to the equilibrium value
of $\gamma$, characterizing the electron-hole superfluid. The plot
in Fig.~\ref{Fig_3D}a represents the low-temperature BCS-like
superfluid for electron-hole pairing with a non-zero  value of the
order parameter $\gamma >0$  at the minimum, while
Fig.~\ref{Fig_3D}b shows the high-temperature non-superfluid EHP for
the zero  value of $\gamma$  at the minimum. Both cases are connected
via the second-order phase transition, as visualized in Fig. \ref{Fig2},
where the normalized free energy $f$ as a function of the order parameter
$\gamma$ is plotted for different values of $\beta u_{0}$ at fixed parameters
$t\beta$ and  $\delta_{0}/2t$.  The curves
for $\beta u_{0} = -50$ and $\beta u_{0} = -60$ demonstrate the
existence of low-temperature BCS-like superfluid with electron-hole
pairing with a non-zero equilibrium value of the order parameter
$\gamma > 0$. The second order phase transition is characterized
by the equilibrium value $\gamma = 0$, which corresponds to the
values of the parameter $\beta u_{0}$ between $\beta u_{0} = -50$
and $\beta u_{0} = -40$, as it is shown in Fig.~\ref{Fig2}. According
to Fig.~\ref{Fig2}, for fixed parameters $\delta_{0}/2t$ and $t\beta$,
the minimal order parameter $\gamma$ increases with decreasing $\beta u_{0}$
if $\beta u_{0} < -50$. The
curves in Fig.~\ref{Fig2} for $\beta u_{0} = -30$ and $\beta u_{0} = -35$
represent the high-temperature non-superfluid EHP with  $\gamma =
0$.

\begin{figure}[tbp]
\psfrag{E1 =0.5}{$\delta_0=t\ \ \ $}
\psfrag{E1 =0}{$\hskip-0.2cm\delta_0=0$}
\psfrag{E1 =1}{$\hskip-0.2cm\delta_0=2t$}
\psfrag{E1 =0.5, V=0}{$\delta_0=t,NP$}
\psfrag{E1 =0, V=0}{$\hskip-0.2cm\delta_0=0,NP$}
\psfrag{E1 =1, V=0}{$\hskip-0.2cm\delta_0=2t,NP$}
\psfrag{beta_c=20}{$\beta=20t$}
\psfrag{beta_c=40}{$\beta=40t$}
\psfrag{beta_c-60}{$\beta=60t$}
\psfrag{Interaction}{Interaction, $u_0/2t$}
\psfrag{Critical Temperature}{Critical temperature, $k_BT_c/2t$}
\psfrag{Interaction}{Interaction, $u_0/2t$}
\psfrag{Chemical Potential}{Chemical potential, $\delta_0/2t$}
\psfrag{Critical Interaction}{Critical interaction, $u_0/2t$}
\includegraphics[width=10.0cm]{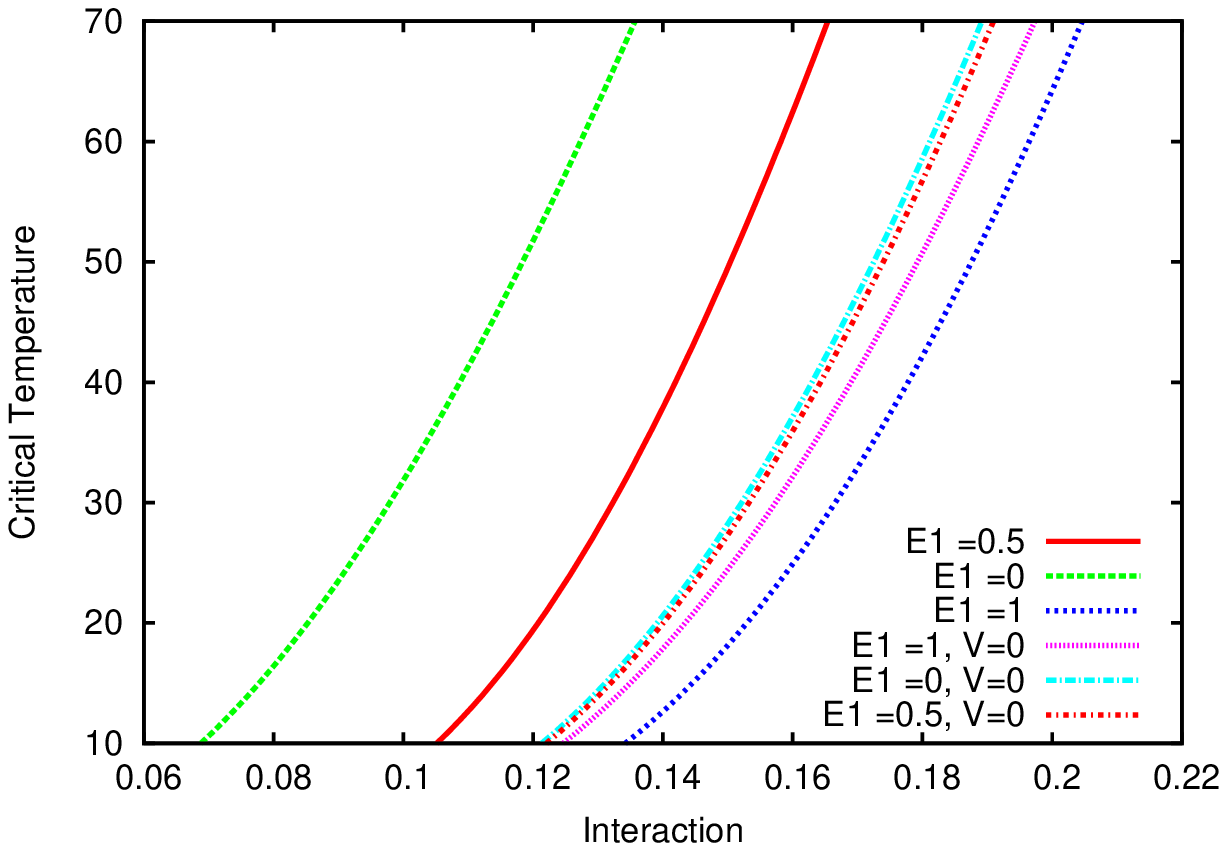}
\includegraphics[width=10.0cm]{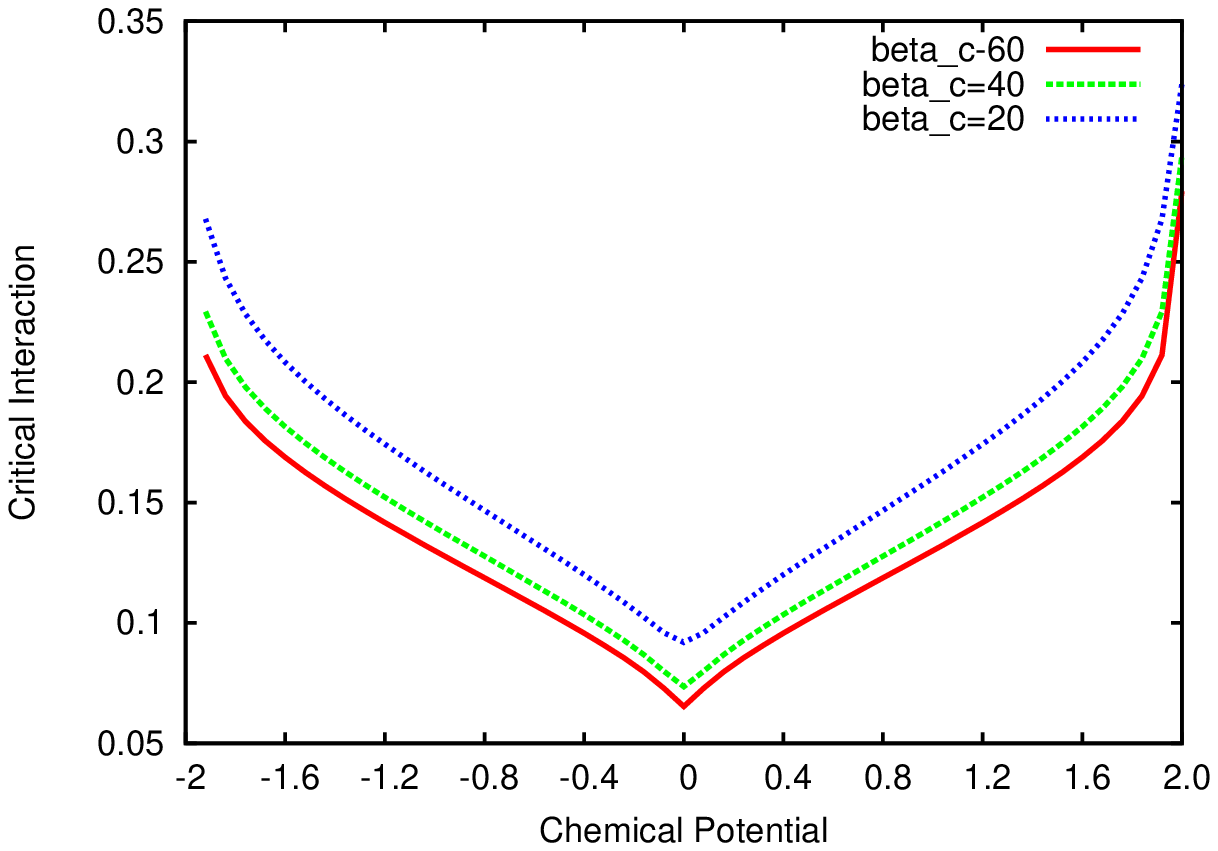}
\caption{
a) Critical temperature as a function of the interaction parameter
for different normalized Fermi energies $\delta_0/2t$.
The curves with index $NP$
are without periodic potential, while the other curves are for a periodic
potential in tight-binding approximation of Eq. (\ref{dispersion}).
\\
b) Critical interaction strength at three fixed inverse temperatures $\beta$ as a function of the
Fermi energy $\delta_0/2t$.
 }
\label{Fig5}
\end{figure}

Fig. \ref{Fig5} can be understood as a $k_BT-u_0$ phase diagram, in which
the different curves indicate the phase boundaries between the
EHP on the left and the superfluid on the right for different values of
the Fermi energy in the case of a parabolic dispersion ($V=0$) and the
case of a periodic potential with dispersion (\ref{dispersion}).
In the presence of the periodic potential the phase boundaries are
more separated for the three values of the Fermi energy $\delta_0=0,0.5t,t$
than those of the parabolic dispersion. On the other hand, all phase
boundaries of the latter occur at smaller values $-u_0$ than those of
periodic potential if $\delta_0=t$.

\section{Discussion and Conclusions}
\label{disc}

In the framework of the mean-field approach for electron-hole
pairing, we applied the tight-binding approximation for the single
electron spectrum of the superlattice created by the external
periodic potential and studied the effect of an additional periodic
potential on the EHP-superfluid transition.
We have assumed for simplicity that the dispersion and the Fermi
surfaces of the electrons and the holes are the same, and
analyzed the phase transition at finite temperatures. 
Our results clearly indicate the possibility to control the
electron-hole superfluid in CQWs or double layers of 2D
material by applying an external periodic potential due to an
attached  periodic gate. An alternative approach is to create a
tunable periodic lattices in two twisted layers of 2D materials
(``magic angle'' bilayers) \cite{tran19}. The analogous effect
occurs in a supersolid, where the crystalline long-range order and
non-crystalline long-range order
co-exist~\cite{Andreev,Chester,Kuklov}. Contrary to a supersolid,
where the crystalline phase is formed due to self-organization, the
band structure in the system under consideration are induced by the
external periodic potential.

A periodic potential creates many bands which are typically separated by
gaps. (Neighboring bands can also touch each other at spectral nodes. This case
though, is not considered here.) To reduce the calculation to a single band we
have assumed that the order parameter $|\Delta|$ is smaller than the gap between the
neighboring band. This allows us to use a single band projection, based on a tight-binding
model. The single band has a lower band edge at energy $E_0$ and an upper band edge
at energy $E_1$, which are the boundaries of integration for the condition of the
critical temperature in Eq. (\ref{mfa_1}).
The $\bk$ integration is determined by the Coulomb interaction. In BCS theory of Cooper pairs
only a small interval around the Fermi energy, whose width is given by the Debye energy
$\hbar\omega_D$, contributes to an attractive interaction:
$E_0=E_F-\hbar\omega_D$, $E_1=E_F+\hbar\omega_D$ \cite{Schrieffer,abrikosov}.
This is different in the excitonic case because the attractive Coulomb interaction exists for all
energies inside the band.

The critical temperature depends on the chemical potential $\delta_0$. Typically it
decreases when $\delta_0$ moves away from zero, as depicted
in Fig. \ref{Fig5}. This is also the case in the absence of the periodic potential which has a
constant density of states (cf. Fig. \ref{Fig5}a).

A one-dimensional periodic potential has an inverse square root singularity at $E=0$, which can
result in an even stronger enhancement of the critical temperature. In that case we must use
the density of states $\rho(E_x)$ for a potential varying in $x$ direction. Then the
condition in Eq. (\ref{mfa_1}) becomes
\beq
\frac{1}{u_0}
=-\frac{1}{p_1-p_0}\int_{p_0}^{p_1}\int_{E_0}^{E_1}\frac{\tanh(\beta_c|2tE_x+p_y^2/2m-\delta_0|)}
{|2tE_x+p_y^2/2m-\delta_0|}\rho(E_x)dE_xdp_y
\ .
\eeq

A critical assumption in our model is that electrons and holes have the same
dispersion. If they had different dispersions we expect a more
complex phase diagram (cf. Ref.\cite{BKZ}). It can even affect the form of the order parameter.
In our study we assume that the pumping beam is circularly polarized,
and hence the excitons are formed only in one of the valleys: $\mathbf{K}$ or
$\mathbf{-K}$~\cite{Xiao,Mak2013}. In this study we address the formation of excitons in one of the 
valleys. Moreover, we can have two electronic species in TMDC materials due to the existence of two
valleys \cite{qiu13}.

It should be a particular interest to
extend our MFA approach to a more complex one with the valley degrees of
freedom and the effective coupling between the two valleys included.
Another interesting extension of the MFA of the present work would
be the inclusion of quantum fluctuations. This would open a wide
avenue for measurements of quantum effects near the EHP-superfluid
transition as well as inside the EHP and the superfluid through
quantum excitations. A first step in this direction was the
calculation of the density-density correlation and the structure
factor, which indicates a characteristic increase near the
transition \cite{BKLZ}. Another possibility is to determine quantum
transport properties in the EHP and the superfluid.

\section*{Acknowledgments}
O.L.B. and R.Ya.K. were supported by US Department of
Defense under Grant No. W911NF1810433. Yu.E.L. was supported by
the Program of Basic Research of HSE and RFBR grants 17-02-01134 and
18-52-00002. K.G.Z. is grateful for support by a grant from the
Julian Schwinger Foundation.

\appendix

\section{Expansion of free energy with respect to the order parameter in
Landau form}

\label{app:A}

From Eq.~(\ref{mfa0}) the dimensionless free energy
$f=-u_{0}F/(k_{B}T)^{2}$ can be written as

\begin{equation}
f=\gamma ^{2}+u_{0}\beta \frac{1}{|B|}\int_{B}\ln \left[ 2\left(
1+\cosh \left( \sqrt{\beta ^{2}\varepsilon _{\mathbf{p}}^{2}+\gamma
^{2}}\right) \right) \right] d^{2}p.  \label{NormalFreeEnergy}
\end{equation}
One can expand the integrant in Eq. (\ref{NormalFreeEnergy}) in
terms of the
power of $\gamma ^{2}$ as%
\begin{eqnarray}
&& \ln \left[ 2\left( 1+\cosh \left( \sqrt{\beta ^{2}\varepsilon _{\mathbf{p}%
}^{2}+\gamma ^{2}}\right) \right) \right] = \ln [2(1+\cosh [\beta
\left\vert \varepsilon _{\mathbf{p}}\right\vert )]+\frac{\tanh
\left( \frac{1}{2}\beta \left\vert \varepsilon
_{\mathbf{p}}\right\vert \right) }{2\beta \left\vert \varepsilon
_{\mathbf{p}}\right\vert }\gamma ^{2}
\nonumber \\
&+& \frac{\beta
^{2}\varepsilon _{\mathbf{p}}^{2}-\beta \left\vert \varepsilon _{\mathbf{p}%
}\right\vert \sinh \left( \beta \left\vert \varepsilon _{\mathbf{p}%
}\right\vert \right) }{8\beta ^{4}\varepsilon
_{\mathbf{p}}^{4}\left(
1+\cosh [\beta \left\vert \varepsilon _{\mathbf{p}}\right\vert ]\right) }%
\gamma ^{4}+...  \label{ExpIntegrant}
\end{eqnarray}
Here to obtain the final expression for the coefficients of the expansion we
have used the following identities:
\begin{equation} 
\sinh 2u=2\sinh u\cosh u,\hspace{1cm}1+\cosh 2u=2\cosh ^{2}u.
\label{Identity}
\end{equation}
Substituting (\ref{ExpIntegrant}) into Eq. (\ref{NormalFreeEnergy})
we present the dimensionless free energy in the Landau
form~\cite{Landau, landau}

\[
f=f_{0}+f_{2}\gamma ^{2}+f_{4}\gamma ^{4}+...,
\]

where

\begin{eqnarray}
f_{0} &=&u_{0}\beta \frac{1}{|B|}\int_{B}\ln [2(1+\cosh [\beta
\left\vert
\varepsilon _{\mathbf{p}}\right\vert )]d^{2}p,  \label{f0} \\
f_{2} &=&1+u_{0}\beta \frac{1}{|B|}\int_{B}\frac{\tanh \left( \frac{1}{2}%
\beta \left\vert \varepsilon _{\mathbf{p}}\right\vert \right)
}{2\beta
\left\vert \varepsilon _{\mathbf{p}}\right\vert }d^{2}p,  \label{f2} \\
f_{4} &=&u_{0}\beta \frac{1}{|B|}\int_{B}\frac{\beta ^{2}\varepsilon _{%
\mathbf{p}}^{2}-\beta \left\vert \varepsilon
_{\mathbf{p}}\right\vert \sinh
\left( \beta \left\vert \varepsilon _{\mathbf{p}}\right\vert \right) }{%
8\beta ^{4}\varepsilon _{\mathbf{p}}^{4}\left( 1+\cosh [\beta
\left\vert \varepsilon _{\mathbf{p}}\right\vert ]\right) }d^{2}p.
\label{f4}
\end{eqnarray}%

\section{Free energy in the case of the periodic potential}

\label{app:B}

The the case of the periodic potential the integration in (\ref{NormalFreeEnergy}) as well as in Eqs.~(\ref{f0}) - (%
\ref{f4}) is taken over the Brillouin zone, implying $|B|$ is the
area of
the Brillouin zone (for the square superlattice of the period $b$: $%
\left\vert B\right\vert =(2\pi \hbar )^{2}/b^{2}$, and, therefore,
the limits of the integration over $p_{x}$ and $p_{y}$ are given by
$-\pi \hbar /b$ and $-\pi \hbar /b$. Assuming that in (\ref{f0}) -
(\ref{f4}) the single-particle energy dispersions versus momentum
for electrons and holes
are the same we can calculate $f_{0},$ $f_{2}$ and
$f_{4}$

\begin{eqnarray}
f_{0} &=&u_{0}\beta \int_{-2}^{2}\frac{\ln [2(1+\cosh [\beta
\left\vert
\delta _{0}-2tE\right\vert )]K\left( \frac{2-\left\vert E\right\vert }{%
2+\left\vert E\right\vert }\right) }{2+\left\vert E\right\vert }dE,
\label{fP0} \\
f_{2} &=&1+u_{0}\beta \int_{-2}^{2}\frac{\tanh \left(
\frac{1}{2}\beta \left\vert \delta _{0}-2tE\right\vert \right)
K\left( \frac{2-\left\vert E\right\vert }{2+\left\vert E\right\vert
}\right) }{2\left\vert \delta
_{0}-2tE\right\vert \times (2+\left\vert E\right\vert )}dE,  \label{fP2} \\
f_{4} &=&u_{0}\beta \int_{-2}^{2}\frac{\beta ^{2}(\delta
_{0}-2tE)^{2}-\beta \left\vert \delta _{0}-2tE\right\vert \sinh
\left( \beta \left\vert \varepsilon _{\mathbf{p}}\right\vert \right)
}{8\beta ^{4}(\delta _{0}-2tE)^{4}\left( 1+\cosh [\beta \left\vert
(\delta _{0}-2tE\right\vert ]\right) }\frac{K\left(
\frac{2-\left\vert E\right\vert }{2+\left\vert E\right\vert }\right)
}{(2+\left\vert E\right\vert )}dE,  \label{fP4}
\end{eqnarray}%
where $K(k)$ is the complete elliptic integral of the first kind.


\begin{thebibliography}{99}


\bibitem{Lozovik} Yu.~E. Lozovik and V.~I. Yudson, Sov. Phys. JETP Lett. {\bf
22}, 26 (1975); Sov. Phys. JETP {\bf 44},  389 (1976).

\bibitem{Shevchenko} S.~I. Shevchenko, Phys. Rev. Lett. {\bf 72}, 3242  (1994).

\bibitem{Littlewood} Xu. Zhu, P.~B. Littlewood, M.~S. Hybertsen and T.~M.
Rice,  Phys. Rev. Lett. {\bf 74},  1633 (1995).

\bibitem{Vignale} S. Conti, G. Vignale and A.~H. MacDonald, Phys. Rev. B {\bf
57},   R6846 (1998).

\bibitem{Ulloa} M.~A. Olivares-Robles and S.~E. Ulloa, Phys. Rev. B  {\bf
64}, 115302 (2001).

\bibitem{DasSarma} D.~S.~L. Abergel, M. Rodriguez-Vega, E. Rossi, and S. Das
Sarma, Phys. Rev. B {\bf 88},  235402 (2013).

\bibitem{Peeters} M. Zarenia, A. Perali, D. Neilson, and F.~M. Peeters, Sci.~Rep.
{\bf 4}, 7319 (2014).

\bibitem{combescot17} M. Combescot, R. Combescot and F. Dubin,
Rep. Prog. Phys. {\bf 80}, 066501  (2017).

\bibitem{Fil} D.~V. Fil and S.~I. Shevchenko, Low~Temp.~Phys. {\bf 44}, 867 (2018).

\bibitem{usp} Yu.~E. Lozovik, Physics-Uspekhi {\bf 188},   1203 (2018).

\bibitem{Perali} A. Perali, D. Neilson, and A.~R. Hamilton, Phys. Rev. Lett.
{\bf 110},   146803 (2013).

\bibitem{Zrenner} T. Fukuzawa, E.~E. Mendez and J.~M. Hong,  Phys. Rev.
Lett. {\bf 64}, 3066 (1990); J.~A. Kash, M. Zachau, E.~E. Mendez,
J.~M. Hong and T. Fukuzawa,  Phys. Rev. Lett. {\bf 66}, 2247 (1991).

\bibitem{Sivan} U. Sivan, P.~M. Solomon and H. Shtrikman, Phys. Rev. Lett.
{\bf 68}, 1196 (1992).

\bibitem{Snoke} S.~A. Moskalenko and D.~W. Snoke, \textit{Bose-Einstein Condensation
of Excitons and Biexcitons and Coherent Nonlinear Optics with
Excitons} (Cambridge University Press, New York, 2000).

\bibitem{Chemla} L.~V. Butov, A. Zrenner, G. Abstreiter, G. Bohm and G.
Weimann, Phys. Rev. Lett. {\bf 73}, 304 (1994); L.~V. Butov, C.~W.
Lai, A.~L. Ivanov, A.~C. Gossard, and D.~S. Chemla, Nature {\bf
417}, 47 (2002); L.~V. Butov, A.~C. Gossard, and D.~S. Chemla,
Nature {\bf 418}, 751 (2002).

\bibitem{Butovr} L.~V. Butov, J. Phys. Condens. Matter {\bf 16},  R1577 (2004).

\bibitem{Timofeev}  A.~V. Larionov and V.~B.
Timofeev,  JETP Lett. {\bf 73}, 301 (2001);  A.~V. Gorbunov and
V.~B. Timofeev, JEPT Lett. {\bf 84}, 329 (2006).

\bibitem{Krivolapchuk} V.~V. Krivolapchuk, E.~S.Moskalenko, and A.~L.
Zhmodikov, Phys. Rev. B {\bf 64}, 045313 (2001).

\bibitem{Snoke_paper} D. Snoke, S. Denev, Y. Liu, L. Pfeiffer, and K. West,
Nature {\bf 418}, 754 (2002).

\bibitem{Snoke_paper_Sc} D. Snoke, Science {\bf 298}, 1368 (2002).

\bibitem{Dubin_PRL} R. Anankine, M. Beian, S. Dang, M. Alloing, E. Cambril, K. Merghem,
C.~G. Carbonell, A. Lematre, and F. Dubin,  Phys. Rev. Lett. {\bf
118},  127402 (2017).

\bibitem{Dean_Nature} J.~I.~A. Li, T. Taniguchi, K. Watanabe, J. Hone, and C.~R.
Dean, Nature Physics  {\bf 13}, 751 (2017).

\bibitem{LB} Yu.~E. Lozovik and O.~L. Berman, JETP~Lett. {\bf 64}, 573 (1996).

\bibitem{BLSC} O.~L. Berman, Yu.~E. Lozovik, D.~W. Snoke, and R.~D. Coalson,
Phys. Rev. B {\bf 70},  235310 (2004).

\bibitem{LBW} Yu.~E. Lozovik and O.~L. Berman, Physica Scripta {\bf
58}, 86 (1998).

\bibitem{As1} G.~E. Astrakharchik, J. Boronat, I.~L. Kurbakov, and Yu.~E.
Lozovik,  Phys.~Rev.~Lett. {\bf 98},  060405 (2007).

\bibitem{As2} A.~E. Golomedov, G.~E. Astrakharchik, and Yu.~E.
Lozovik, Phys.~Rev.~A {\bf 84}, 033615 (2011).

\bibitem{JBD} Y.~N. Joglekar, A.~V. Balatsky, and S. Das Sarma, Phys. Rev. B
{\bf 74}, 233302  (2006).

\bibitem{Snoke_review} D.~W. Snoke, in \textit{Quantum Gases: Finite Temperature and
Non-Equilibrium Dynamics} (Vol. 1, Cold Atoms Series), N.~P.
Proukakis, S.~A. Gardiner, M.~J. Davis, and M.~H. Szymanska, eds.,
Imperial College Press, London, 2013.

\bibitem{Pepper_review} K. Das Gupta, A.~F. Croxall, J. Waldie, C.~A.
Nicoll, H.~E. Beere, I. Farrer, D. A. Ritchie, and M. Pepper, Advances in
Condensed Matter Physics, Volume 2011, Article ID 727958.

\bibitem{Butov_JPCM} L.~V. Butov, J. Phys.: Condens. Matter {\bf 19}, 295202 (2007).

\bibitem{Kukushkin} V.~V. Solov'ev, I.~V. Kukushkin, J. Smet,
K.~von~Klitzing, and W. Dietsche, JETP Letters \textbf{83}, 553 (2006).

\bibitem{Dubin} M. Alloing, M. Beian, M. Lewenstein, D. Fuster, Y. Gonz\'{a}%
lez, L. Gonz\'{a}lez, R. Combescot, M. Combescot, and F. Dubin, Europhys.
Lett. \textbf{107}, 10012 (2014).

\bibitem{Rapaport} K. Cohen, Y. Shilo, K. West, L. Pfeiffer, and R.
Rapaport, Nano~Lett.  {\bf 16}, 3726 (2016).

\bibitem{Butov_per} M. Remeika, J.~C. Graves, A.~T. Hammack, A.~D. Meyertholen, M.~M. Fogler, L.~V. Butov, M. Hanson, and A.~C.
Gossard,    Phys.~Rev.~Lett. {\bf 102}, 186803 (2009).

\bibitem{BLG} O.~L. Berman, Yu.~E. Lozovik, and G. Gumbs, Phys.~Rev.~B {\bf
77}, 155433 (2008).

\bibitem{Sokolik} Yu.~E. Lozovik and A.~A. Sokolik, JETP Lett. {\bf 87}, 55 (2008);
  Phys. Lett. A {\bf 374}, 326 (2009).

\bibitem{Bist} R. Bistritzer and A.~H. MacDonald, Phys. Rev. Lett. {\bf 101}, 256406 (2008).

\bibitem{BKZg} O.~L. Berman, R.~Ya. Kezerashvili, and K. Ziegler, Phys.~Rev.~B {\bf 85}, 035418 (2012).

\bibitem{Efimkin} D.~K. Efimkin, Yu.~E. Lozovik, and A.~A. Sokolik,  Phys.~Rev.~B {\bf 86}, 115436 (2012).

\bibitem{EM} J.~P. Eisenstein and A.~H. MacDonald, Nature (London) {\bf
432}, 691 (2004).

\bibitem{Kormanyos} A. Korm\'{a}nyos, G. Burkard, M. Gmitra, J. Fabian, V. Z%
\'{o}lyomi, N.~D. Drummond, and V. Fal'ko, 2D Mater. \textbf{2},
022001 (2015).

\bibitem{Mak2010} K.~F. Mak, C. Lee, J. Hone, J. Shan, and T.~F. Heinz,
Phys.~Rev.~Lett.  \textbf{105}, 136805 (2010).


\bibitem{Xiao} D. Xiao, G.~B. Liu, W. Feng, X. Xu, and W. Yao, Phys.~Rev.~Lett. {\bf 108}, 196802 (2012).


\bibitem{Mak2013} K.~F. Mak,  K. He, C. Lee, G.~H. Lee, J. Hone, T.~F. Heinz, and J. Shan, Nat. Mater. {\bf
12}, 207 (2013).

\bibitem{Fogler} M.~M. Fogler, L.~V. Butov, and K.~S. Novoselov, Nature Commun. {\bf 5},
4555 (2014).

\bibitem{Calman} E.~V. Calman, C.~J. Dorow, M.~M. Fogler, L.~V. Butov, S.
Hu, A. Mishchenko, and A.~K. Geim,  Appl.~Phys.~Lett. {\bf 108},
101901 (2016).

\bibitem{MacDonald_TMDC} F.-C. Wu, F. Xue, and A.~H. MacDonald, Phys.~Rev.~B {\bf 92}, 165121
(2015).

\bibitem{BK}   O.~L. Berman and R.~Ya. Kezerashvili, \prb {\bf 93}, 245410 (2016).

\bibitem{BK2}   O.~L. Berman and R.~Ya. Kezerashvili, \prb {\bf 96}, 094502 (2017).

\bibitem{Schrieffer} J.~R. Schrieffer, \textit{Theory of Superconductivity} (New
York: Benjamin, 1964).

\bibitem{DeGennes} P.~G. De Gennes, \textit{Superconductivity of Metals and Alloys}
(W. A. Benjamin, 1966).

\bibitem{BKZ} O.~L. Berman, R.~Ya. Kezerashvili, and K. Ziegler,
Physica E {\bf 71}, 7 (2015).


\bibitem{Landau} L.D. Landau and E.M.~Lifshitz, \textit{Statistical
Physics, Part 1} (Pergamon Press; 3rd edition, Oxford, NY, 1980).


\bibitem{landau}
E.M. Lifshitz and L.P. Pitaevskii, {\it Statistical Physics}
(Pergamon Press, Oxford, 1980) Pt. 2.

\bibitem{lee18}
H. Lee, K. Paeng and I.S. Kim, Synthetic Metals {\bf 244}, 36 (2018).

\bibitem{morell10}
E. Suárez Morell et al, Phys. Rev. B {\bf 82}, 121407 (2010).

\bibitem{bistritzer11}
R. Bistritzer and A.H. MacDonald, Proc. Natl. Acad. Sci. USA {\bf 108}, 12233 (2011).

\bibitem{Ziman} J.~M. Ziman, \textit{Principles of the Theory of Solids}
(Cambridge University Press; 2nd edition, Cambridge, 1979).

\bibitem{Simon} S.~H. Simon, \textit{The Oxford Solid State Basics} (Oxford
University Press, Oxford, 2013).

\bibitem{Ashcroft} N.~W. Ashcroft and N.~D. Mermin, \textit{Solid State
Physics} (Sounders College, New York, 1976).

\bibitem{gonis}
A. Gonis, {\it Green functions for ordered and disordered systems}
(North-Holland, Amsterdam, 1992).

\bibitem{tran19}
K. Tran et al., Nature {\bf 567}, 71 (2019).

\bibitem{Andreev} A.~F. Andreev and I.~M. Lifshitz, Sov. Phys. JETP \textbf{%
29}, 1107 (1969).

\bibitem{Chester} G.~V. Chester, \pra {\bf 2}, 256 (1970).

\bibitem{Kuklov} M. Boninsegni, A.~B. Kuklov, L. Pollet, N.~V. Prokof'ev,
B.~V. Svistunov, and M. Troyer, \prl {\bf 97}, 080401 (2006).

\bibitem{abrikosov}
A.A. Abrikosov, L.P. Gorkov, and I.E. Dzyaloshinski, {\it Methods of
quantum field theory in statistical physics}, Dover Publications
(New York 1975).


\bibitem{qiu13}
D.Y. Qiu, F.H. da Jornada, and S.G. Louie, Phys. Rev. Lett. {\bf 111}, 216805 (2013).




\bibitem{BKLZ}
O.L. Berman, R.Ya. Kezerashvili, Yu.~E. Lozovik, K. Ziegler, Physica
E {\bf 92}, 1 (2017).




\end{thebibliography}
\end{document}